\documentclass[apj]{emulateapj}
\usepackage{apjfonts}
\usepackage{amsmath}
\usepackage{natbib}

\submitted{Accepted by the Astrophysical Journal Letters.}

\usepackage{graphicx}
\usepackage[figuresright]{rotating} 
\usepackage{float}       
\usepackage{afterpage}   

\newcommand {\AN}{annihilation}
\newcommand {\SO}{SN 1006}
\newcommand {\FLUX}{photons cm$^{-2}$ s$^{-1}$}
\newcommand {\GR}{$\gamma$-ray}
\newcommand{\integral}{\emph{INTEGRAL}}
\newcommand{\wsim}{\ensuremath{\sim}}

\begin{document}


\title{Searching for annihilation radiation from SN 1006 with SPI on \emph{INTEGRAL}}

\author{E. Kalemci\altaffilmark{1,2},
        S. E. Boggs\altaffilmark{1,3},
        P. A. Milne \altaffilmark{4},
        S. P. Reynolds\altaffilmark{5}
}

\altaffiltext{1}{Space Sciences Laboratory, 7 Gauss Way, University of
California, Berkeley, CA, 94720-7450, USA.}

\altaffiltext{2}{Current address: Sabanc\i\ University, Orhanl\i\
-Tuzla 34956, \.Istanbul, Turkey.}

\altaffiltext{3}{Department of Physics, University of California,
366 Le Conte Hall, Berkeley, CA, 94720-7300, USA.}

\altaffiltext{4}{Steward Observatory, 933 N. Cherry Ave., Tucson,
AZ, 85721, USA.}

\altaffiltext{5}{Department of Physics, NC State University, 2700
Stinson Drive, Box 8202, Raleigh, NC 27695, USA.}


\begin{abstract}

Historical Type Ia supernovae are a leading candidate for the source
of positrons observed through their diffuse annihilation emission in
the Galaxy. However, search for annihilation emission from
individual Type Ia supernovae has not been possible before the
improved sensitivity of \integral. The total 511 keV annihilation
flux from individual SNe Ia, as well as their contribution to the
overall diffuse emission, depends critically on the escape fraction
of positrons produced in $^{56}$Co decays. Late optical light curves
suggest that this fraction may be as high as 5\%. We searched for
positron \AN\ radiation from the historical Type Ia supernova \SO\
using the SPI instrument on \integral. We did not detect significant
511 keV line emission, with a 3$\sigma$ flux upper limit of 0.59
$\times$ 10$^{-4}$ \FLUX\ for \wsim 1 Msec exposure time, assuming a
FWHM of 2.5 keV. This upper limit corresponds to a 7.5\% escape
fraction, 50\% higher than the expected 5\% escape scenario, and
rules out the possibility that Type Ia supernovae produce all of the
positrons in the Galaxy (\wsim 12\% escape fraction), if the mean
positron lifetime is less than 10$^{5}$ years. Future observations
with \integral\ will provide stronger limits on the escape fraction
of positrons, the mean positron lifetime, and the contribution of
Type Ia supernovae to the overall positron content of the Galaxy.

\end{abstract}




\section{Introduction}\label{sec:intro}

Searching for \GR\ lines from supernovae (SNe) remains one of the primary
goals of \GR\ astrophysics, and also \integral\ \citep{WinklerC03}, as \GR\
 line studies can probe the nucleosynthesis and explosion kinematics of
these events. Many of the nuclear decay chains ($^{56}$Co,
$^{44}$Ti) that produce \GR\ lines in young SNe also produce
positrons, making them a prime candidate for the source of positrons
for the diffuse Galactic annihilation emission.  However, there has
not been a detection of positron \AN\ radiation from an individual
SN or a SN remnant (SNR). The main uncertainties on the expected
\AN\ fluxes from young SNe are the mean lifetime and escape fraction
of positrons. For SNe Type Ia, the lifetime of a positron will be
small in the initial SN, but may become as high as $> 10^{4}$ years
as the SN expands. The mean positron lifetime will be even longer,
$> 10^{5}$ years, in cases where the thermalization takes place in
the interstellar medium (ISM), rather than in the ejecta
\citep{Guessoum91}. If the ejecta's magnetic field is weak, and/or
radially combed, then 95\% of the $^{56}$Co decay positrons
annihilate promptly, and 5\% escape to annihilate on a longer
time-scale \citep{Chan93}. Alternatively, a tangled and strong
magnetic field would confine \wsim 100\% of the $^{56}$Co positrons
resulting prompt annihilation, leaving a much smaller \AN\ radiation
flux from the delayed $^{44}$Ti decays ($\tau$ \wsim 85 years).

Simulations of late optical B \& V light curves of Type Ia SNe
indicate that the 5\% escape of the $^{56}$Co positrons is the more
probable scenario, assuming the optical light curves trace the
bolometric luminosity \citep{Milne99}. A later study also showed
that 16 of 22 SNe Ia observed at late epochs exhibit the same shape
BVRI light curves as the initial sampling of 10 SNe Ia, strengthening
 the conclusion of 5\% escape fraction \citep{Milne01}. However,
 excess NIR emission at late epochs was reported recently from
 SN~1998bu and SN 2000cx \citep{Spyromilio04,Sollerman04}. This excess
 may be due to emission shifting into the NIR wavelength range from
 the optical range. The optical-NIR bolometric light curves for those
 SNe favor full positron trapping. Observations of positron annihilation
  radiation from individual SNRs can improve our understanding of
  magnetic field configurations in these objects, as well as determine
  their role in producing the diffuse Galactic \AN\ emission.

The positrons are initially hot, and slow down by Coulomb losses,
ionization, and excitation. Once they are cool enough (less than a
few hundred eV), the positrons can have charge exchange with
neutrals, recombine radiatively with free electrons, and/or
annihilate directly with free or bound electrons, producing a line
at 511 keV, and a positronium continuum
\citep{Bussard79,Guessoum05}. The fraction of the photons that are
emitted in the line and the continuum depends on the properties of
the annihilating medium. The recent analysis of the SPI data yields
positronium fractions close to 0.95 for the diffuse Galactic \AN\
radiation \citep{Churazov05, Jean05}. For the width of the line, SPI
finds 2.4$\pm$0.3 FWHM, if the line is approximated as a simple
Gaussian \citep{Churazov05}.\footnote{Recent analysis of the SPI
observations of the galactic center region suggest that the line may
be composed of a narrow (FWHM $\sim$ 1.3 keV) and broad (FWHM $\sim$
5.1 keV) component \citep{Jean05}.} Note that the positronium
fraction and the FWHM values for the Galaxy are not necessarily
appropriate for individual SNe \AN s. The positronium fraction
depends on the temperature and the ionization state of the medium
\citep{Guessoum91,Jean05}, and the FWHM can be broadened by Doppler
effects if the annihilation takes place in the ejecta.

\begin{figure}
\plotone{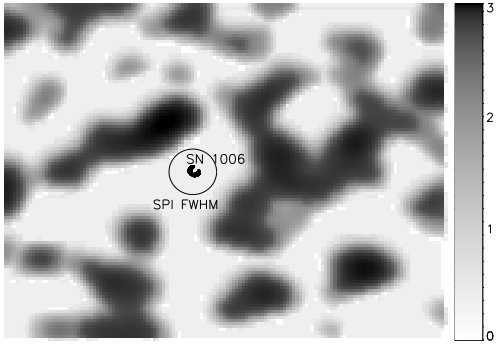} \caption{\label{fig:spifig} SPI significance image
in 508.5--513.5 band of the fully coded field of view. The
\emph{ASCA} contours of \SO\ is overlaid to indicate the extent and
position of the source. The 3$^{\circ}$ FWHM of the imaging
resolution of SPI is shown for comparison. The significance scale is
plotted on the right. }
\end{figure}

For weak and/or radially combed magnetic fields that permit 5\%
escape, positron transport simulations estimate that roughly 8
$\times$ 10$^{52}$ positrons escape from a given SN Ia
\citep{Milne99}. When combined with the rate of 0.5 Type Ia SNe per
century, a steady state flux of 9 $\times$ 10$^{-4}$ \FLUX\ would be
expected if the positrons are assumed to be annihilated at a
distance of 8 kpc (distance to the Galactic center). The latest
measurements of SPI yields a total Galactic 511 keV flux of 1.5--2.9
$\times$ 10$^{-3}$ \FLUX\ depending on the Galactic distribution
model \citep{Knodlseder05}. The higher end of that flux range is in
agreement with OSSE on \emph{CGRO}, \emph{SMM}, and \emph{TGRS}
observations of 511 keV line emission \citep{Milne01P}. The expected
flux from the 5\% escape case is therefore 30\% -- 60\% of the total
flux, with the higher fluxes producing the lower SN fractions.
Therefore, if positrons escape the ejecta, but are confined to local
regions of the SNe, the maps of 511 should trace the recent SN Ia
history. The bulge-to-disk flux ratio is confined to be within the
range of 1--3 (and luminosity ratio of 3--9). This ratio is
consistent with no Galactic scale diffusion, and makes Type Ia SNe
prime candidates for the source of Galactic bulge positrons, along
with low mass X-ray binaries (see \citealt{Knodlseder05} for a
thorough discussion of potential candidates for diffuse \AN\
radiation).

The most promising individual source candidate to search for a
positron \AN\ line is \SO, a recent SN thought to be of Type Ia. Our
group observed \SO\ with \integral\ for \wsim 1000 ks, with the main
goals of characterizing the hard X-ray emission using ISGRI and
JEM-X instruments \citep{Kalemci05_sub}, and searching for 511 keV
emission line with SPI. In this letter we discuss the
 results of the analysis of the SPI data, and place limits on the 511 keV line
and positronium continuum emission from \SO.


\section{Observations, Analysis and Results}\label{sec:obs}

SPI is a coded-aperture telescope using an array of 19 cooled germanium
detectors for high-resolution spectroscopy \citep{Vedrenne03}. It works in
20 keV -- 8 MeV band, and has an energy resolution of \wsim 2 keV at 511 keV.
The fully coded field of view is 16$^{\circ}$, and the angular resolution
is 3$^{\circ}$.

The \integral\ observations of \SO\ took place in two sets. The first 250 ks
set was conducted early in the mission, between MJD~52650 and MJD~52659
corresponding to \integral\ revolutions 30 and 32. These observations will be
denoted as "Set I". The second 750 ks set was conducted between MJD~53024 and
MJD~53034 during revolutions 155-158, which will be denoted as "Set II". For
SPI, Set I has all 19 detectors working, whereas for Set II, the active
detectors were reduced to 18 after the loss of Detector 2.

Before any analysis, we filtered out the pointings with high
Anti-Coincidence Shield rates, mostly occurring during the entry and
exit of the radiation belts. We used OSA 4.2, SPIROS 6.1, and
single-detector events for the analysis. Several background models
were tried with ``GEDSAT'' as the main model, which assumes that the
background level is proportional to the product of saturated
detector trigger rate and live-time. We used the continuum energy
band of 523-545 keV for normalization. We tried normalization with
an OFF observation (empty field observation in Revolution 130), and
also using a template file (provided by J. Kn\"odlseder). The
details of the background methods and information about the template
file can be found in \cite{Knodlseder05}. Finally, we tried the mean
count modulation method in SPIROS which does not require a prior
background determination. Maximum likelihood method in SPIROS was
used for imaging. Note that \SO\ is \wsim 30$^{\prime}$ in diameter,
and is effectively a point source for SPI. Set I and Set II were
analyzed separately. We used 5, 7, and 10 keV energy intervals
encompassing 511 keV to factor in possibilities for both narrow and
broad emission. We also searched for the positronium continuum in
200--500 keV band with SPI.

Fig.~\ref{fig:spifig} shows the SPIROS sigma image in the energy band of
508.5--513.5 keV, using the mean count modulation method. We obtained similar
images in larger energy bands, and different background methods. We have not
detected significant 511 keV line emission in any of the energy band intervals
around 511 keV. The positronium component was not detected either.
Table~\ref{table:res} shows the details of the search and 3$\sigma$
sensitivity limits of SPI for the given energy band and observing time.
The combined upper limit is for 18 detectors in Set I and Set II.


\begin{table}[ht]
\caption{\label{table:res} SPI Upper Limits}
\begin{tabular}{lc|c|c} \hline \hline
 511 keV Line & Set I (250 ks)& Set II (750 ks) & Combined \\ \hline
FWHM (keV) & Flux\footnote{All fluxes are 3$\sigma$ upper limits, ph cm$^{-2}$ s$^{-1}$} & Flux  & Flux \\
2.5  & 1.15 $\times$ 10$^{-4}$ & 0.70 $\times$ 10$^{-4}$ &  0.59 $\times$ 10$^{-4}$ \\
3.5  & 1.35 $\times$ 10$^{-4}$ & 0.81 $\times$ 10$^{-4}$ &  0.70 $\times$ 10$^{-4}$ \\
5.0  & 1.66 $\times$ 10$^{-4}$ & 0.96 $\times$ 10$^{-4}$ &  0.83 $\times$ 10$^{-4}$ \\
\hline Positronium Continuum & & \\ \hline
350--500 keV band & 5.8 $\times$ 10$^{-4}$ & 3.4 $\times$ 10$^{-4}$ &  3.0 $\times$ 10$^{-4}$\\
\end{tabular}
\end{table}


\section{Discussion}\label{sec:discussion}

\SO\ is an ideal candidate to search for 511 keV positron \AN\ line
emission, as it is a well studied, historical Type Ia SN with
relatively well determined characteristics. The distance is \wsim
2.2 kpc, the age is 999 years, the angular size is \wsim
30$^{\prime}$, and the ejecta's current electron density is 0.7
cm$^{-3}$ \citep{Winkler03,Long03,MilneD71,Milne99,Reynolds98}.


\begin{figure}
\plotone{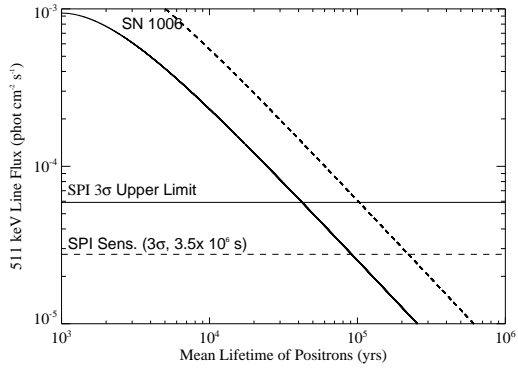} \caption{\label{fig:spipred} Predicted 511 keV
line fluxes from \SO\ as a function of positron mean lifetime. The
solid and dotted curves are for 5\% and 12\% escape fractions,
respectively. The expected SPI (5$\sigma$) sensitivity to the 511
keV line for 3500 ksec, and the current SPI 3$\sigma$ upper limits
for the emission from the SNRs are shown for reference. }
\end{figure}

Predictions of the current 511 keV line and  positronium continuum
fluxes from \SO\ depend upon understanding the complicated
interaction of the SN shock with the ISM, and the resulting degree
of magnetic confinement of positrons \citep{Ruiz97}. Simulations
have shown that positrons that escape from the SN ejecta typically
have $\sim$ 500 keV of energy upon their escape
\citep{Chan93,Milne99}. The range of 500 keV positrons in the ISM is
such that few positrons would be expected to thermalize within 1000
years \citep{Guessoum91}. This situation is even worse at the
location of \SO, where the ISM density appears even lower than four
media treated by \cite{Guessoum91}. Thus, if the shock fails to
confine escaping positrons, perhaps even accelerating the positrons,
then \SO\ would be only a faint source of annihilation radiation,
even if positrons escaped the ejecta. The primary argument against
positrons crossing the shock in this fashion are circumstantial, as
the distribution of annihilation radiation from SNe would then trace
the distribution of matter in the Galaxy. It has been shown that the
diffuse 511 keV emission mapping results from OSSE and SPI
\citep{Purcell97, Milne01, Knodlseder05, Teegarden05} indicate a
strong bulge component, with a bulge-to-disk luminosity ratio of
3--9. By contrast, the matter content of the Galaxy is concentrated
in the disk \citep[bulge-to-disk mass ratio is only
0.3--1.0,][]{Robin03}, which appears to eliminate large-scale
positron transport from source to annihilation site (not just from
SNe, but in general).

If the magnetic field in the shock successfully confines the
positrons, they will repeatedly traverse the inside of the bubble,
being reflected numerous times. Preliminary simulations of this
scenario suggest that the positrons will encounter a sufficient
amount of matter to be thermalized and annihilate on a 1000 year
timescale. However, these simulations are too crude for us to
confidently predict an annihilation rate at 1000 years after the SN
event. The more detailed simulation will be conducted for the
analysis of the entire dataset including this and the near-future
observations.

In Fig.~\ref{fig:spipred}, we show the predicted 511 keV fluxes from
\SO\ as a function of mean positron lifetime for the 5\% escape
fraction. The figure also shows the current 3$\sigma$ SPI upper
limit, and the 3$\sigma$ sensitivity level that will be achieved
after the approved future observations (2.5 Msec). If 5\% of the
positrons indeed escape the ejecta and are locally confined with a
lifetime less than $10^5$ years, then SPI is expected to detect 511
keV annihilation emission from SN 1006 once the additional
observations are performed.

Fig.~\ref{fig:spipred} also shows the predicted fluxes for \SO\ if
the positron yield is increased by a factor of 2.5, so that SNe Ia
could account for {\bf{all}} Galactic positrons. The SPI
non-detection suggests against this possibility if the mean positron
lifetime is less than 10$^{5}$ years. Note that, in case of no
escape from the remnant, the expected 511 keV flux from only
$^{44}$Ti decays is no larger than 4 $\times$ 10$^{-6}$ \FLUX\ from
\SO, which is far too low to be detectable by SPI.

We stress that the conclusions above depend strongly on the
assumption of no escape to the ISM, and thermalization in the
ejecta. The future observations of \SO, and other SN Ia remnants of
different ages (both younger and older), will provide essential
information for understanding positron transport in SNRs. A
non-detection would still leave an ambiguity; the reason can either
be \wsim 0\% escape fraction from the ejecta, or very long
thermalization timescale of positrons in the ISM. On the other hand,
detection of \AN\ radiation would be a strong indicator of
thermalization in the ejecta, and confirm whether SNe Ia are
important contributors of galactic positrons.


\acknowledgments E.K. is supported by the European Commission
through the FP 6 Marie-Curie International Reintegration Grant
(INDAM), and also acknowledges partial support of T\"UB\.ITAK. E.K.
and S.E.B. acknowledge NASA grants NAG5-13142 and NAG5-13093. S.P.R.
acknowledges support from NASA grant NAG5-13092. E. K. thanks Pierre
Jean for useful discussions, and also thanks the reviewer Richard
Lingenfelter for his remarks which significantly improved the
letter.








\end{document}